\documentclass[authoryear]{FLO_v1}%

\usepackage{graphicx}
\usepackage{upgreek}
\usepackage{amsmath,amssymb,amsfonts}
\usepackage{mathrsfs}
\usepackage{amsthm}
\usepackage[figuresright]{rotating}
\usepackage{appendix}
\usepackage[authoryear]{natbib}
\usepackage{ifpdf}
\usepackage[T1]{fontenc}
\usepackage{newtxtext}
\usepackage{newtxmath}
\usepackage{textcomp}
\usepackage{xcolor}

\usepackage[colorlinks,allcolors=blue]{hyperref}
\definecolor{jourcolor}{cmyk}{1,0.57,0.01,0.38}
\hypersetup{
    colorlinks,%
    citecolor=jourcolor,%
    filecolor=jourcolor,%
    linkcolor=jourcolor,%
    urlcolor=jourcolor
}

\usepackage{physics}
\newcommand\Rey{\mbox{\textit{Re}}}  

\reversemarginpar
\usepackage[color=red!80!black, textcolor=white, textsize=small]{todonotes}

\usepackage[locale=UK,
number-mode=math,
unit-mode=text,
per-mode=symbol,
number-unit-product=\ ,
retain-zero-exponent=false]{siunitx}[=v2]
\DeclareSIUnit{\pixel}{px}
\DeclareSIUnit{\fps}{fps}

\usepackage{soul}
\usepackage[nameinlink,noabbrev]{cleveref}%
\crefname{figure}{Figure}{Figures}
\crefname{equation}{Equation}{Equations}

\newcommand{\kindex}[2]{\ensuremath{{#1}_{\scalebox{0.6}{#2}}}}

\articletype{PREPRINT}

\begin{document}

\title{Lagrangian analysis of bio-inspired vortex ring formation}
\author{Mrudhula Baskaran, Karen Mulleners$^{\ast}$}%
\address{Institute of Mechanical Engineering, École polytechnique fédérale de Lausanne (EPFL), Lausanne, Switzerland}
\corres{*}{Corresponding author. E-mail:\emaillink{karen.mulleners@epfl.ch}}


\abstract{
Pulsatile jet propulsion is a highly energy-efficient swimming mode used by various species of aquatic animals that continues to inspire engineers of underwater vehicles. Here, we present a bio-inspired jet propulsor that combines the flexible hull of a jellyfish with the compression motion of a scallop to create individual vortex rings for thrust generation. Similar to the biological jetters, our propulsor generates a non-linear time-varying exit velocity profile and has a finite volume capacity. The formation process of the vortices generated by this jet profile is analysed using time-resolved velocity field measurements. The transient development of the vortex properties is characterised based on the evolution of ridges in the finite-time Lyapunov exponent field and on local extrema in the pressure field derived from the velocity data. Special attention is directed toward the vortex merging observed in the trailing shear layer. During vortex merging, the Lagrangian vortex boundaries first contract in the stream-wise direction before expanding in the normal direction to keep the non-dimensional energy at its minimum value in agreement with the Kelvin-Benjamin variational principle. The circulation, diameter, and translational velocity of the vortex increase due to merging. The vortex merging takes place because the velocity of the trailing vortex is higher than the velocity of the main vortex ring prior to merging. The comparison of the temporal evolution of the Lagrangian vortex boundaries and the pressure based vortex delimiters confirms that features in the pressure field serve as accurate and robust observables for the vortex formation process.
}

\maketitle

\begin{boxtext}
\textbf{\mathversion{bold}Impact Statement}
Aquatic animals that utilise jet propulsion intuitively manipulate the exit velocity profile and nozzle diameter in time to adapt their swimming performance.
To integrate this capability in human-engineered underwater vehicles, detailed knowledge about the influence of non-linear time-varying velocity profiles on the development and formation number of vortex rings is required.
We have designed a bio-inspired device that ejects propulsive vortex rings by compressing a polymer bulb to experimentally study the formation process of a vortex ring with a time-varying jet profile.
Our findings reveal insight into the evolution of the integral vortex quantities for these generalised conditions and contribute to the fundamental understanding of vortex merging in the wake of a vortex ring.
The comparison of the temporal evolution of the Lagrangian vortex boundaries and the pressure based vortex delimiters confirms that features in the pressure field serve as accurate and robust observables for the vortex formation process.
These results disclose new opportunities to use local pressure sensors as input for closed loop control of the temporal exit velocity profile to improve the propulsion efficiency of under water vehicles utilising pulsatile jet propulsion.
\end{boxtext}

\section{\label{sec:intro}Introduction}
Underwater vehicles are used for a wide range of applications, such as marine exploration \citep{bayatEnvironmentalMonitoringUsing2017}, ecosystem monitoring \citep{whitcombAdvancesUnderwaterRobot2000}, and ocean cleaning \citep{zahugiOilSpillCleaning2013}.
One class of underwater vehicles relies on propellor-based technology and turbomachinery for long-range propulsion.
These vehicles are not suitable for application in sensitive or fragile environments like coral reefs, or in constrained environments with limited space for manoeuvering \citep{mohseniPulsatileVortexGenerators2006a}.
Another class of underwater vehicles consists of bio-inspired swimmers, which mimic the kinematics of biological organisms like fish, jellyfish, and cephalopods for locomotion \citep{zhuTunaRoboticsHighfrequency2019, weymouthUltrafastEscapeManeuver2015a, robertsonRoboScallopBivalveInspired2019a}.
Bio-inspired vehicles are highly controllable and can explore a wide range of environments.

One of the mechanisms commonly exploited by bio-inspired vehicles for transport is pulsatile jet propulsion, which relies on the periodic ejection of vortex rings to create thrust \citep{kruegerSignificanceVortexRing2003, whittleseyOptimalVortexFormation2013a}.
The canonical device used to generate and study vortex rings is the piston cylinder apparatus.
A translating piston pushes fluid from within a cylinder through an opening or nozzle at the end of the cylinder where a shear layer forms and rolls-up into a coherent ring vortex \citep{dabiriOptimalVortexFormation2009}.
The vortex ring grows larger as more fluid is ejected by the piston.
However, the growth of the vortex does not continue indefinitely.
Beyond a limiting non-dimensional vortex formation number, additional fluid supplied by the vortex generator is no longer directly entrained by the vortex ring and instead forms a trailing shear layer \citep{gharibUniversalTimeScale1998}.

A vortex ring produced by a classical piston cylinder apparatus with a constant piston velocity simultaneously attains its maximum circulation, reaches its minimum non-dimensional energy, and outpaces its feeding shear layer after approximately four convective time scales or stroke ratios \citep{gharibUniversalTimeScale1998}.
This limiting value of the vortex formation number is not universal and depends on the geometry of the outlet nozzle or orifice \citep{limbourgExtensionUniversalTime2021, dabiriStartingFlowNozzles2005, kriegNewKinematicCriterion2021,ofarrellPinchoffNonaxisymmetricVortex2014a}, the presence of a uniform background co- or counterflow \citep{dabiriDelayVortexRing2004a, kruegerFormationNumberVortex2006a}, and the temporal evolution of the piston velocity \citep{rosenfeldCirculationFormationNumber1998, zhaoEffectsTrailingJet2000, shusserEffectTimedependentPiston2006, olcayMomentumEvolutionEjected2010a}.
Different time-dependent profiles of the exit velocity can noticeably alter the formation number of the ring vortex, without significantly affecting its circulation.
Aquatic animals that utilise jet propulsion intuitively manipulate the exit velocity profile and nozzle diameter to adapt their swimming performance \citep{dabiriFastswimmingHydromedusaeExploit2006,lipinskiFlowStructuresFluid2009,gemmellCoolYourJets2021}.
The practical desire to integrate this capability in human-engineered underwater vehicles reopens fundamental questions on vortex ring formation.
Detailed knowledge of the influence of arbitrary time-varying velocity profiles on the formation number and the pinch-off of vortex rings are crucial to improve the controllability and energy-efficiency of underwater vehicles that operate using jet propulsion.

An objective and frame-independent method to identify vortex pinch-off and characterise the formation process is based on Lagrangian coherent structures \citep{hallerLagrangianStructuresRate2001}.
Lagrangian coherent structures are maxima, or ridges, in the positive and negative finite-time Lyapunov exponent (FTLE) fields.
They demarcate the vortex ring and separate regions in the flow that are dynamically different \citep{ofarrellLagrangianApproachIdentifying2010, shaddenLagrangianAnalysisFluid2006a}.
The FTLE ridges emerge between the vortex ring and the trailing shear layer when the vortex no longer accepts additional vorticity and pinches off \citep{shaddenLagrangianAnalysisFluid2006a}.
Alternatively, prominent features such as local extrema in the pressure field can serve as instantaneous indicators of vortex pinch-off.
Regions of elevated pressure in front and behind the vortex ring are called leading and trailing pressure maxima.
These maxima indicate regions where vorticity not longer passes into the vortex ring similar to the positive and negative FTLE ridges \citep{lawsonFormationTurbulentVortex2013}.
The emergence of the trailing pressure maximum coincides with the formation number of the vortex ring and is a necessary condition for pinch-off \citep{schlueter-kuckPressureEvolutionShear2016}.

Here, we present a bio-inspired vortex generator as a potential propulsion mechanism for underwater vehicles.
The generator consists of a soft elastic bulb that is compressed by two rigid arms.
Similar to biological jetters, our vortex generator produces a non-linear time-varying exit velocity profile and has a finite volume capacity which limits the maximum stroke length that the device can attain more than in classical piston cylinder arrangements.
It bears similarities with synthetic jet actuators \citep{Shuster.2007, lawsonFormationTurbulentVortex2013, VanBuren.2014, Straccia.2020}.
We experimentally study the vortex ring formation using time-resolved velocity field measurements.
The transient development of the vortex characteristics are analysed based on the evolution of ridges in the finite-time Lyapunov exponent field and on local extrema in the pressure field derived from the velocity data.
Special attention is directed toward the vortex merging event observed in the trailing shear layer.
The robustness of the emergence of pressure maxima as observables to identify the end of the vortex formation process is also evaluated.
The findings will aid the further design and control of underwater vehicles that operate using pulsatile jet propulsion.

\section{\label{sec:materials}Materials and methods}
\subsection{Vortex generator}
\begin{figure}[b!]
\includegraphics{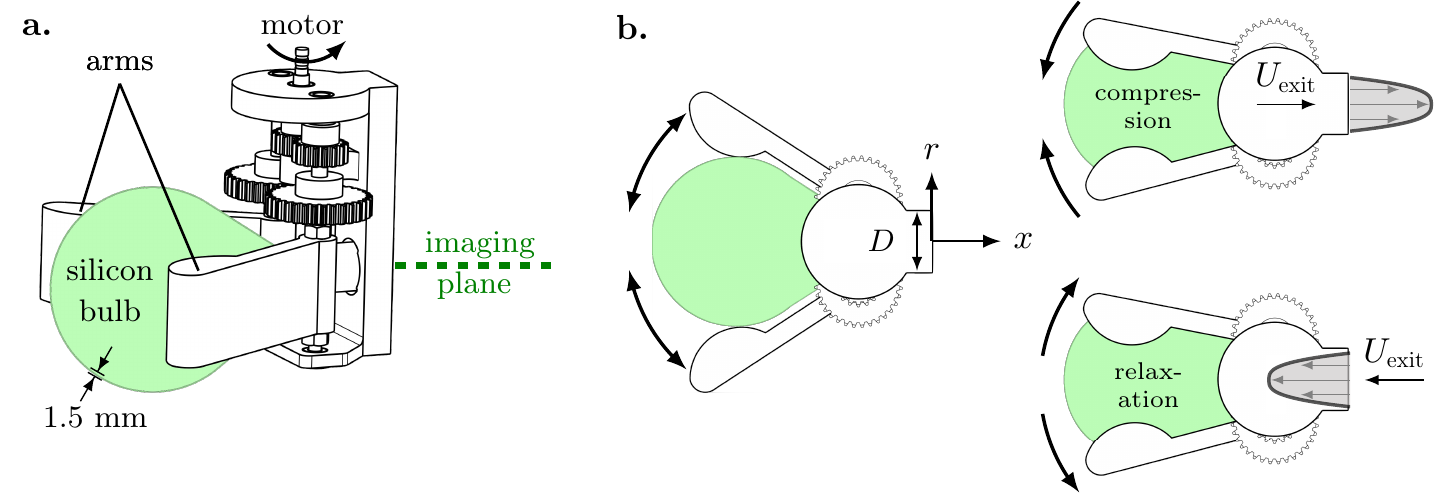}
\caption{(\textbf{a.}) Schematic of the vortex generator including the gear system to translate the rotation on the main axis into a symmetric compression of the bulb by the two arms.
(\textbf{b.}) Relaxed and compressed states of the bulb.}\label{fig:scallop}
\end{figure}

The bio-inspired vortex generator designed for this study combines the flexible bell of a jellyfish \citep{weymouthUltrafastEscapeManeuver2015a, xuSquidinspiredRobotsPerform2021} with the kinematics of bivalves mollusks or scallops \citep{robertsonRoboScallopBivalveInspired2019a}.
The generator presented in \cref{fig:scallop} produces vortex rings by compressing an elastic bulb with two rigid arms and ejecting fluid through a circular nozzle with diameter $D=\SI{14}{\milli\meter}$.
The silicon bulb is prepared by mixing a silicone moulding compound (Zhermack Elite Double 32 shore A base) and a catalyst in a 1:1 ratio and centrifuging the mixture.
The mixture is then poured into a plastic mould prescribing the shape of the bulb and rotated slowly for \SI{30}{minutes} to ensure a homogeneous thickness of \SI{1.5}{\milli\meter} across the bulb.
The bulb has a volume of \SI{150}{\milli\liter} when uncompressed (\cref{fig:scallop}b).
The body and arms of the vortex generator are 3D printed with standard clear resin using a stereolithographic printer (Formlabs Form 2).
The motion of both arms is controlled by a brushless servo motor (Maxon EC-max) along a single motor shaft via the use of a gear box, which ensures symmetrical compression of the bulb (\cref{fig:scallop}a).
Commands to the motor are sent via a motor controller (DMC-4040, Galil Motion Control, USA).

\subsection{Experimental setup}
The vortex generator is placed in a rectangular glass tank filled with water.
Time-resolved particle image velocimetry (PIV) is used to measure the velocity field in a stream-wise symmetry plane of the vortex.
Polystyrene particles with a diameter of \SI{56}{\micro\meter} are used as seeding particles.
The starting position of the arms is the angle at which they touch the bulb without compressing it (\cref{fig:scallop}b).
The angle of the arms is varied with a sinusoidal profile, to smoothly compress and relax the bulb with a prescribed amplitude and time period.
The bulb walls are thick enough such that the material does not stretch upon mechanical compression by the arms.
The arms mechanically limit the volume the bulb can contain.
When the arm are opened, the bulb returns to its original form.

Two high-power light-emitting diodes (LED) (LED Pulsed System, ILA\_5150 GmbH, Germany) create a light sheet in the horizontal stream-wise plane cutting through the centre of the exit nozzle (\cref{fig:scallop}a).
The applicability of these high-power LED for PIV has been demonstrated previously by \citet{buchmannPulsedHighpowerLED2012, krishnaFlowfieldForceEvolution2018}.
The LED are operated in continuous mode during image acquisition.
A mirror angled at \SI{45}{\degree} is placed underneath the tank and a high-speed camera (Photron Fastcam SA-X2) records \SI{1024x512}{px} images with an acquisition rate of \SI{2000}{\hertz}.
The start of the PIV image acquisition coincides with the start of the compression of the vortex generator.
Consecutive particle images are correlated using a multigrid evaluation method with an initial window size of \SI{32x32}{px} and step size of \SI{3}{px}, or \SI{90}{\percent} overlap.
This overlap was optimal for not artificially smoothing the velocity gradient fields \citep{inproceedings, kindlerAperiodicityFieldFullscale2011}.
This corresponds to a physical grid spacing of \SI{0.21}{\milli\meter} or \SI{0.015}{D}.
The magnification and extent of the spatial domain were selected to ensure that the full formation process of the vortex ring could be observed by a single camera.
The high temporal resolution comes at the expense of a lower spatial resolution of the vortex rings.
Lagrangian vortex analysis methods are used in combination with more classical Eulerian methods to compensate for the lower spatial resolution and exploit the information available through the high temporal resolution.

\subsection{Finite-time Lyapunov exponent field and ridge computation}
The candidate material boundaries of the generated ring vortex are identified from ridges in the finite-time Lyapunov exponent field (FTLE).
The FTLE fields are calculated directly from the measured time-resolved velocity fields by artificially seeding and convecting fluid particles forward or backward in time to obtain the forward or positive pFTLE and backward or negative nFTLE fields.
The initial position of the fluid particle is indicated by $\vb{x}$ and their positions after a given integration time $\kindex{T}{f}$ are found by advecting the particles with the flow using a fourth order Adam-Bashforth-Moulton integration scheme.
The flow map $\kindex{\phi}{t}^{t+\kindex{T}{f}}$ describes the displacement of the particles between time $t$ and $t+\kindex{T}{f}$.
The spatial gradient of the flow map gives us the Cauchy-Green strain tensor whose largest eigenvalue (\kindex{\lambda}{max}) is referred to as the coefficient of expansion $\sigma^{\kindex{T}{f}}(\vb{x},t)$ \citep{greenUsingHyperbolicLagrangian2010a}.
The scalar FTLE field is then defined as \citep{hallerLagrangianCoherentStructures2002}:
\begin{equation}
\textrm{FTLE}^{\kindex{T}{f}}(\vb{x},t) = \frac{1}{2\kindex{T}{f}} \ln\left(\kindex{\lambda}{max}\left( \left[\grad \kindex{\phi}{t}^{t+\kindex{T}{f}} \right]^* \left[\grad\kindex{\phi}{t}^{t+\kindex{T}{f}}\right] \right) \right) = \frac{1}{2\kindex{T}{f}} \ln\left(\sigma^{\kindex{T}{f}}(\vb{x},t)\right)
\end{equation}
where $^*$ is the matrix transpose operator.

Candidate vortex boundaries manifest as ridges or maxima of the scalar FTLE field \citep{hallerLagrangianCoherentStructures2002, shaddenLagrangianAnalysisFluid2006a, greenDetectionLagrangianCoherent2007a, greenUsingHyperbolicLagrangian2010a}.
Ridges in the positive FTLE field are candidate repelling material lines and correspond to regions where there is a maximum divergence of fluid particle trajectories over time.
Ridges in the negative FTLE field are candidate attracting material lines and correspond to regions where there is maximum attraction of fluid particle trajectories over time.
The ridges are computed here using a ridge tracking algorithm, similar to the one described by \cite{lipinskiRidgeTrackingAlgorithm2010a}.
The algorithm locates grid points with maximum intensity and performs a search within the adjacent grid points to determine the next point on the ridge.
An adjacent grid point is selected as the next ridge point if it has a similar or larger magnitude of the finite-time Lyapunov exponent than the current grid point.
The integration time \kindex{T}{f} used for the data presented here is \SI{0.4}{\second} or \SI{80}{\percent} of the full compression-relaxation cycle.

\section{\label{sec:results}Results and discussion}
\subsection{Fluid ejection and entrainment}
\begin{figure}[tb!]
\includegraphics{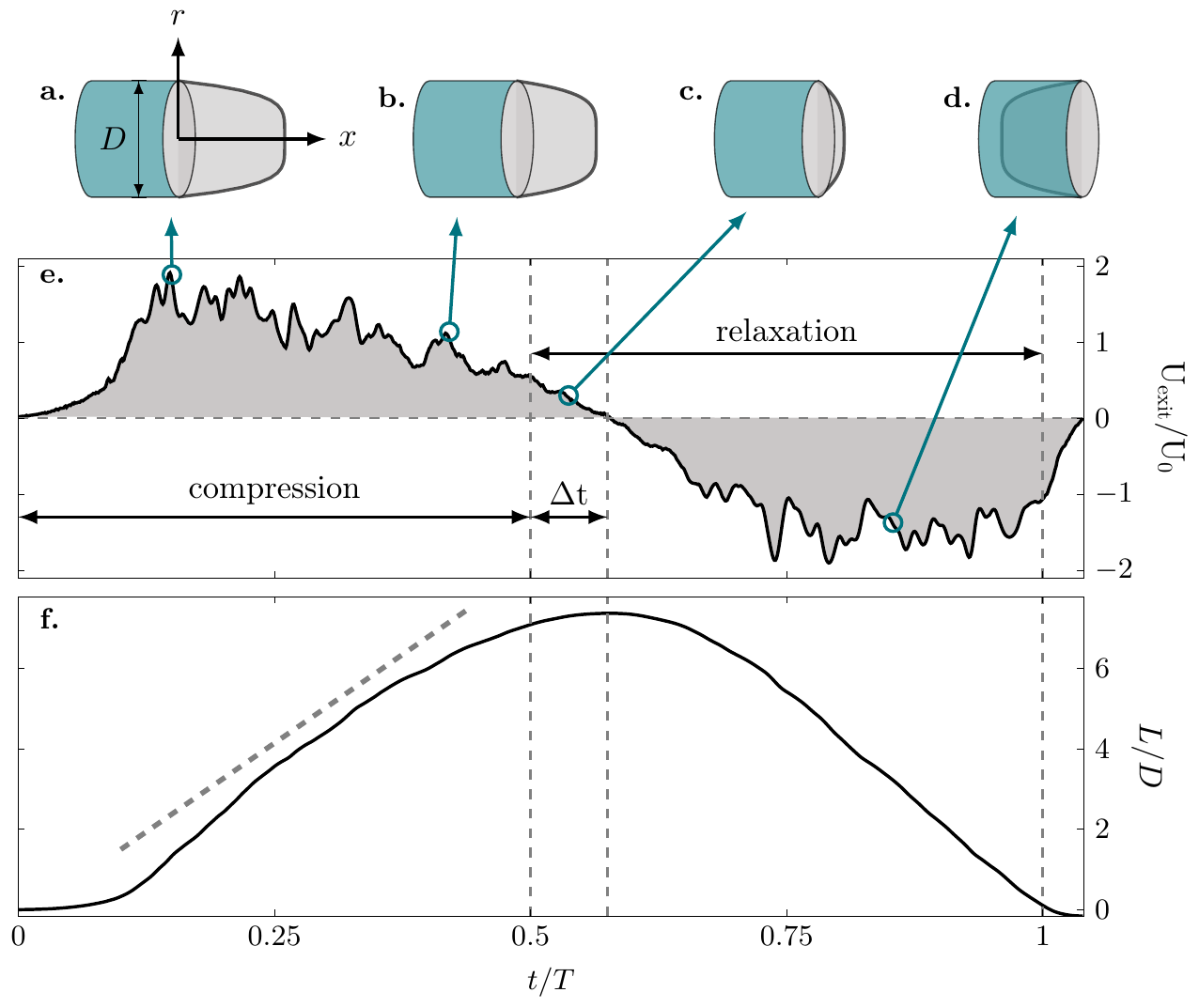}
\caption{(\textbf{a})-(\textbf{d}) Reference cylindrical coordinate system and schematics of the velocity profiles at the nozzle exit at different times during the compression-relaxation cycle.
Fluid leaving the nozzle moves in the positive $x$-direction which is referred to as the stream-wise direction.
(\textbf{e}) Temporal evolution of the measured diameter averaged exit velocity \kindex{U}{exit} normalised by the characteristic velocity \kindex{U}{0} over the duration of one compression-relaxation cycle $T$.
(\textbf{f}) Non-dimensional stroke ratio obtained from integrating \kindex{U}{exit} in time.}\label{fig:uf}
\end{figure}

Our vortex generator operates by compressing and relaxing its deformable bulb.
Compression ejects fluid into the wake and relaxation entrains fluid by allowing the bulb to return to its original form.
The temporal evolution of the average velocity of the fluid that is ejected or entrained by the propulsor is obtained by integrating the vertical velocity profile directly behind the nozzle exit of the propulsor.
This average exit velocity across the nozzle diameter $D$ is defined in cylindrical coordinates as
\begin{equation}
\label{equation:uf}
\kindex{U}{exit}(t) = \frac{4}{\pi D^2} \int\limits_{0}^{2\pi}\int\limits_{0}^{D/2} u(r,x=0,t)\, r \, \dd r \dd\theta 
\end{equation}
where $u(r,x=0,t)$ is the stream-wise velocity component directly behind the nozzle exit at $x/D = 0$.
Schematic representations of the velocity profiles at the nozzle exit and the temporal evolution of the averaged exit velocity are presented in \cref{fig:uf} over the duration of one compression-relaxation cycle $T$.
Positive and negative values of the exit velocity $\kindex{U}{exit}$ correspond respectively to fluid ejection and fluid entrainment.

The exit velocity $\kindex{U}{exit}$ in \cref{fig:uf} is normalised by $\kindex{U}{0}$, the characteristic velocity of the system, which is defined as
\begin{equation}
\kindex{U}{0} = \frac{8\kindex{V}{0}}{\pi D^2 T}\quad.
\label{equation:estar}
\end{equation}
Here, $\kindex{V}{0}$ is the total volume of fluid ejected from the propulsor during the duration of bulb compression, $T/2$.
The velocity \kindex{U}{0} is the velocity of the equivalent constant uniform flow that yields the same ejected volume of fluid over the time of bulb compression.
The velocity \kindex{U}{0} is \SI{415}{\milli\meter\per\second} for the data presented here, and the corresponding Reynolds number \kindex{\Rey}{D} based on the nozzle exit diameter and \kindex{U}{0} is \num{5820}.

Due to the particular design of our vortex generator, the exit velocity increases slowly when the compression starts, but rapidly catches up and reaches a maximum value of $\kindex{U}{exit}/\kindex{U}{0}\approx 1.9$ at $t/T=0.13$.
At $t/T = 0.5$, the kinematics of the propulsor cause a transition from bulb compression to bulb relaxation.
The fluid does not does respond directly to the transition of the kinematics and continues to flow out of the bulb for $\Delta t/T\approx \num{0.07}$ after bulb relaxation has begun.
A shorter lag in the response of the flow with respect to the forcing kinematics is present at the end of cycle.
The flow continues to be entrained for $\Delta t/T\approx \num{0.04}$ after the driving kinematics have stopped.
The slight asymmetry in the fluid ejection-entrainment response to the symmetric compression-relaxation motion is characteristic of this particular driving kinematics and could be compensated for or enhanced by using more complex kinematics.

The length of an equivalent cylindrical column of fluid ejected by the vortex propulsor during compression and relaxation is called the stroke length.
It is defined here as:
\begin{equation}
\label{equation:ld}
L(t) = \int\limits_{0}^{t} \kindex{U}{exit} (\tau)\, \dd\tau\quad.
\end{equation}
The stroke length is analogous to the distance travelled by the piston in a piston cylinder apparatus to eject the same volume of fluid.
The temporal evolution of the stroke to diameter ratio $L/D$ is presented in \cref{fig:uf}f.
After the initial slow response of the flow to the compression kinematic, the stroke ratio increases non-linearly during the compression phase for $\num{0}<t/T<\num{0.57}$.
The stroke ratio attains a maximum value of \num{7} around $t/T=0.57$ and decreases approximately linearly with a rate $\dd L/\dd t= \num{-5.7} D/T $ during the relaxation phase for $\num{0.57}<t/T<\num{1.04}$.

The temporal velocity profile of the fluid ejected by the vortex generator is not constant and leads to a non-linear variation of the stroke ratio.
This allows us to characterise the formation process of a vortex ring generated by a non-linear evolution of the stroke ratio, which has not yet received much attention.
In addition to the nonlinear fluid ejection profile, the fluid entrainment process is unique to the design of our propulsor, and enables the periodic ejection of multiple vortex rings.
The effect of the time-varying exit velocity profile and the fluid entrainment process on the temporal evolution of the propulsive force will be the subject of future investigations.
Here, we focus solely on the flow field created by our propulsor.
Our goal is to characterise the formation process of a vortex ring generated by the arbitrary fluid ejection profile and identify observable quantities that can aid a future optimisation of similar robotic devices that utilise pulsatile jet propulsion.

\begin{figure}[tb!]
\includegraphics{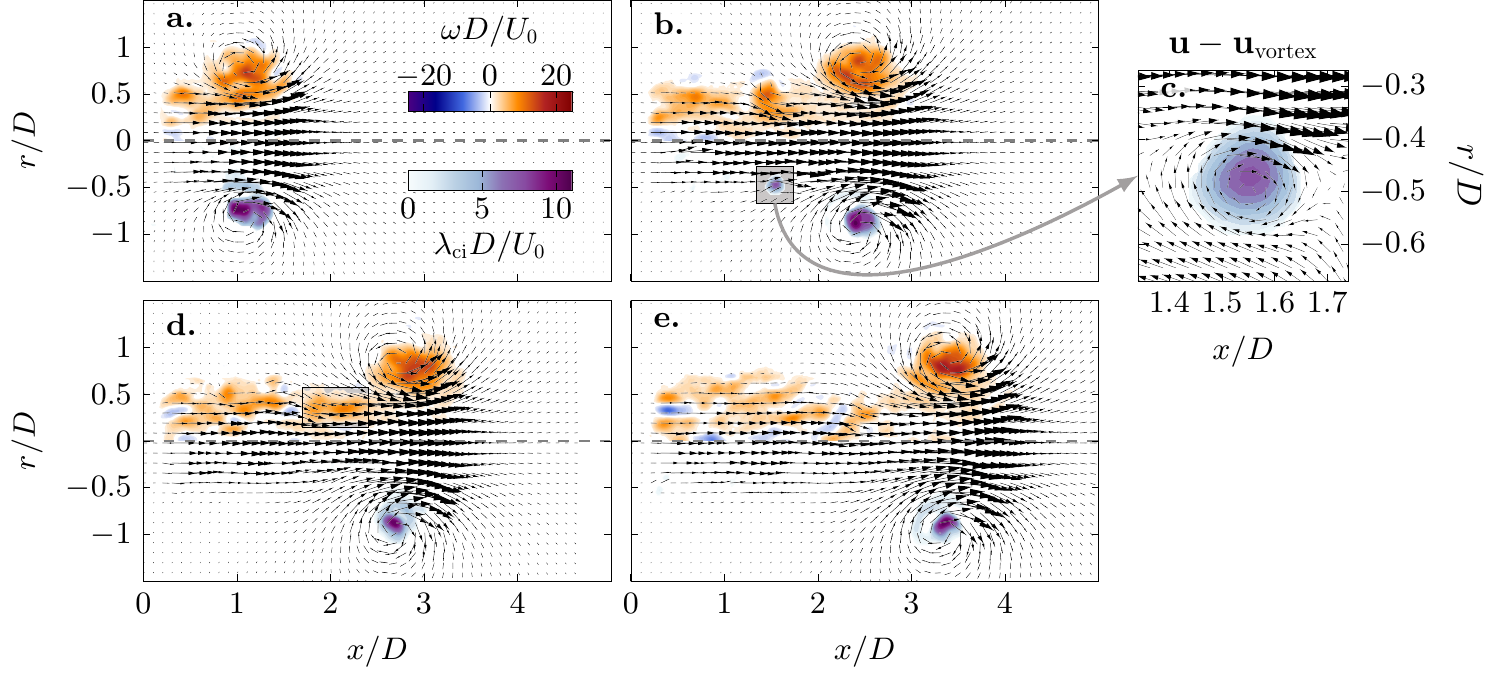}
\caption{Snapshots of experimentally observed vortex ring development during the compression phase of the vortex generator at
(\textbf{a})  $L/D$ = \num{3.4}, (\textbf{b},\textbf{c}) $L/D$ = \num{5.0}, (\textbf{d}) $L/D$ = \num{5.5}, and (\textbf{e}) $L/D$ = \num{5.9}.
The colours in the top half of the snapshots indicate values of the experimentally measured out-of-plane vorticity component $\omega$.
The colours in the bottom half of the snapshots indicate the experimentally obtained swirling strength $\kindex{\lambda}{ci}$.
(\textbf{c}) Zoomed-in view on a secondary vortex in the trailing shear layer with velocity vectors indicating the relative velocity with respect to the velocity measured in the centre of the secondary vortex.
The vorticity associated with the secondary vortex in (\textbf{d}) is indicated by the small box.
}\label{fig:vort}
\end{figure}

\subsection{Vortex formation process}
When fluid is ejected by our vortex propulsor a coherent vortex ring is formed.
The growth of the vortex ring is presented by instantaneous snapshots of the flow field in \cref{fig:vort}.
The arrows represent the two dimensional velocity field $\vb{u}=(u,v)$ in the measurement plane.
The out-of-plane vorticity component, $\omega$, is computed from the in-plane velocity field and is show in the top half of the snapshots in \cref{fig:vort}.
The colours in the bottom half of the snapshots indicate the swirling strength, $\kindex{\lambda}{ci}$, which is the imaginary part of the complex eigenvalues of the velocity gradient tensor \citep{zhouMechanismsGeneratingCoherent1999}.
The swirling strength is a robust indicator of vortices in shear layers where high concentrations of vorticity exist and obfuscate the presence or absence of vortices.

The shear layer that emerges at the boundary between the ejected fluid and the surrounding quiescent flow immediately rolls-up into a vortex ring.
The coherent core is indicated by localised region of non-zero values of the swirling strength (\cref{fig:vort}a).
The vortex ring rapidly convects away from the nozzle exit while flow is still being ejected and a trailing shear layer appears between the vortex ring and the nozzle exit (\cref{fig:vort}b).
At $L/D=\num{5}$, a secondary or trailing vortex is present in the trailing shear layer.
Based on the vorticity concentration alone, it is difficult to distinguish between a strong shear flow and rotation, but the swirling strength distribution is conclusive (\cref{fig:vort}b,c).
The velocity vectors in the zoomed-in view of the secondary vortex in \cref{fig:vort}c represent the velocity relative to the velocity vector measured in the centre of the secondary vortex ($\vb{u}-\kindex{\vb{u}}{vortex}$).
The centre of the vortex is identified as the location of maximum swirling strength.
This secondary vortex core is not persistent and is no longer visible in the swirling strength field for $L/D>\num{5.5}$ when the primary vortex travels further downstream and moves away from the trailing shear layer (\cref{fig:vort}d-e).
The vorticity associated with the secondary vortex is still present and has moved closer to the primary vortex (indicated by the box in \cref{fig:vort}d) suggesting that the primary and secondary vortices begin merging.

\begin{figure}[tb!]
\includegraphics{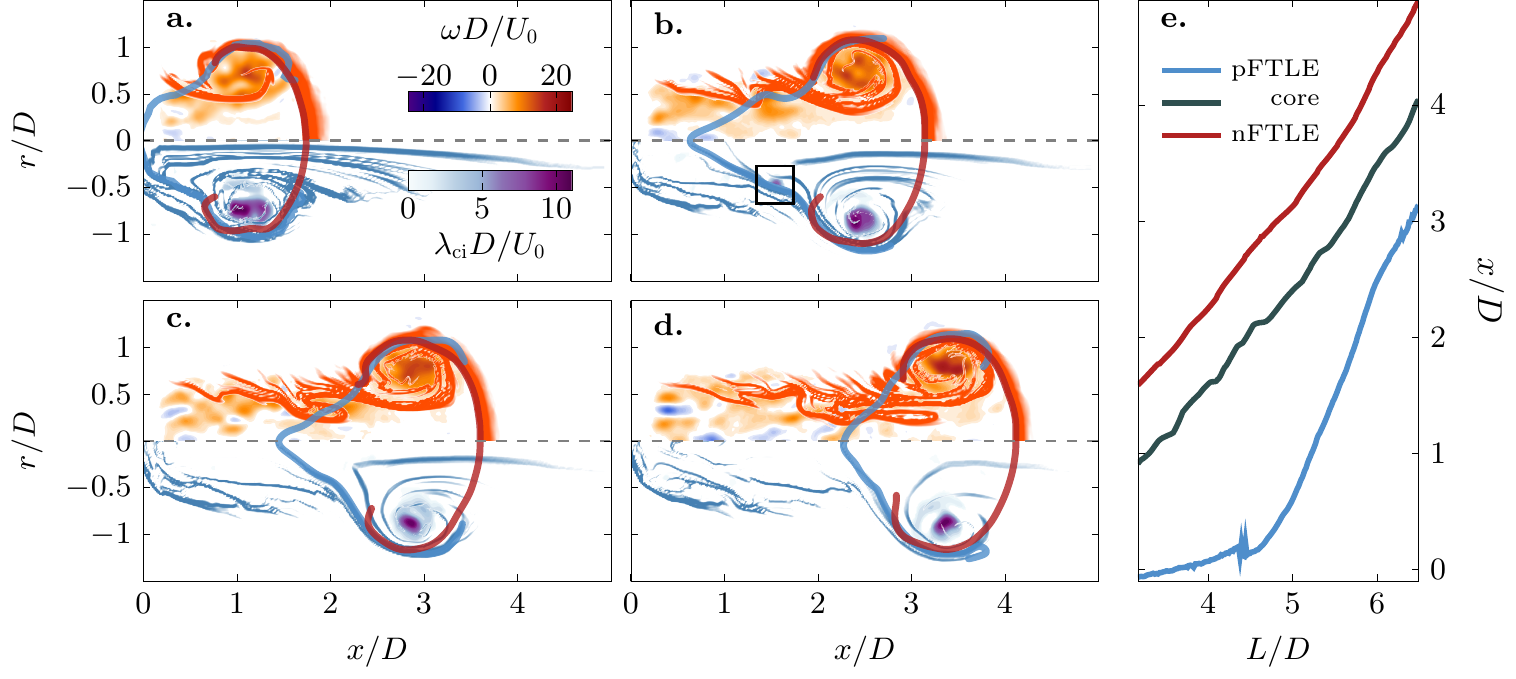}
\caption{Positive and negative finite-time Lyapunov exponent (FTLE) ridges, computed from experimental data, indicating the boundaries of the vortex ring presented atop the vorticity and $\kindex{\lambda}{ci}$ fields for (\textbf{a}) $L/D$ = \num{3.4}, (\textbf{b}) $L/D$ = \num{5.0}, (\textbf{c}) $L/D$ = \num{5.5}, and (\textbf{d}) $L/D$ = \num{5.9}.
(\textbf{e}) Evolution of the stream-wise location of the vortex core and of the crossing of the positive and negative FTLE ridges with the centreline as a function of the stroke ratio.
} \label{fig:ftle}
\end{figure}

To confirm the merging of the primary and secondary vortices, we analyse the temporal evolution of the Lagrangian coherent structures in the positive and negative finite-time Lyapunov exponent (FTLE) fields, which mark the boundaries of the vortex.
\Cref{fig:ftle} shows the positive and negative FTLE ridges and fields atop the vorticity and swirling strength fields.
The positive ridge is the upstream boundary of the vortex ring along which particle trajectories are attracted.
The negative ridge is the downstream boundary of the vortex ring from which particle trajectories are repelled.
The two ridges can be identified once the core of the primary vortex ring has moved one nozzle diameter away from the exit.
The evolution of the stream-wise location of the vortex core and of the intersections of the positive and negative FTLE ridges with the centreline are presented as a function of the stroke ratio $L/D$ in \cref{fig:ftle}e.
As the vortex convects away from the nozzle exit, the positive FTLE ridge lags behind the vortex core (\cref{fig:ftle}b,e).
The positive FTLE ridge encloses the secondary vortex core indicated by the box in \cref{fig:ftle}b.
For $L/D>5$ the positive FTLE ridge accelerates and catches up with the vortex core, pushing the vorticity associated with the secondary vortex core to merge with the vorticity of the primary core.
The distance between the positive FTLE ridge and the core reaches a steady state at $L/D \approx 6$ such that the Lagrangian boundaries symmetrically enclose the vortex core (\cref{fig:ftle}d).

The evolution of the positive ridge during vortex merging enables the mixing of two dynamically different regions of fluid: fluid inside the primary vortex ring and fluid that is in the trailing shear layer including the secondary vortex ring \citep{ofarrellLagrangianApproachIdentifying2010}.
At the end of merging, the positive and negative FTLE ridges are symmetric with respect to the vortex core and the vortex ring has separated from the trailing shear layer.
A symmetric definition of the vortex ring by the FTLE ridges matches the intuitive idea of a pinched-off vortex ring, with the ridges acting as physical barriers that prevent the entrance of additional vorticity into the ring.

\subsection{Evolution of vortex topology during vortex merging}
\begin{figure}[tb!]
\includegraphics{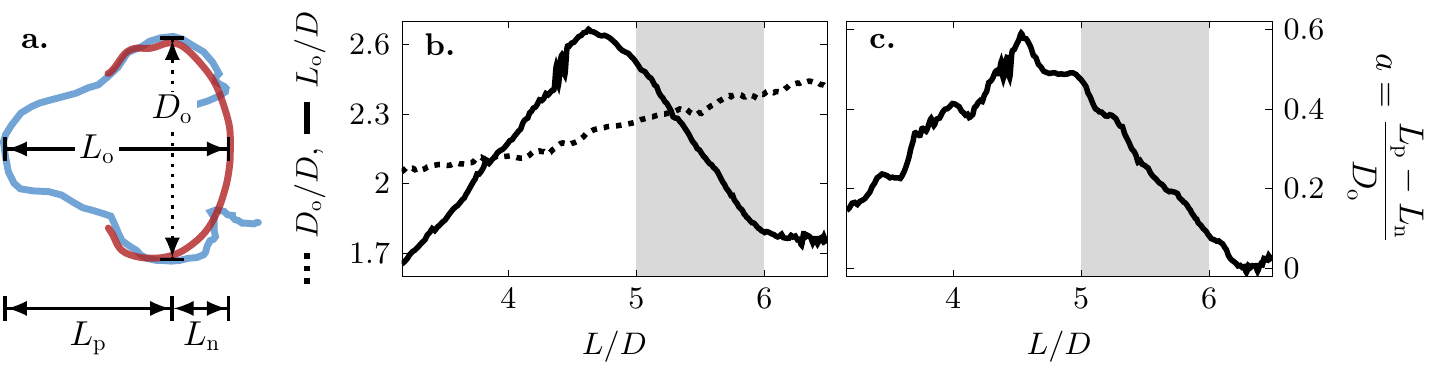}
\caption{Evolution of the shape and asymmetry of the experimentally observed vortex ring.
(\textbf{a}) Vortex boundaries indicated by the FTLE ridges including the definition for the vortex stream-wise length and outer diameter.
Temporal evolution of the (\textbf{b}) vortex length and outer-diameter and (\textbf{c}) the asymmetry parameter $a$ defined by \cref{eq:asymm} as a measure for the degree of asymmetry of the FTLE boundaries with respect to the vortex core.
The shaded area indicates when vortex merging occurs.
}\label{fig:viscues}
\end{figure}

The distance between the intersections of the positive and negative FTLE ridges with the centreline is used to define the stream-wise length, \kindex{L}{o}, of the vortex.
The largest distance in the radial direction between the topmost and bottommost points on the positive FTLE ridge is used here to define the outer diameter, \kindex{D}{o}, or height of the vortex.
The definition of these vortex shape characteristics are indicated in \cref{fig:viscues}a and their temporal evolution for $L/D>3$ is presented in \cref{fig:viscues}b.
For lower stroke ratios, we cannot yet identify the downstream FTLE ridge to determine the vortex length.

Around $L/D=3$, the vortex outer diameter is about twice the nozzle diameter and larger than the stream-wise length.
The outer diameter increases approximately linearly during the rest of the compression phase of the bulb to $\kindex{D}{o} = \SI{2.4}{D}$ at $L/D=6.5$.
The stream-wise length increases faster than the outer diameter as the positive FTLE ridge lags behind and reaches a maximum value of $\kindex{L}{o}=\SI{2.65}{D}$ at $L/D \approx 4.6$, yielding a minimum aspect ratio of $\kindex{L}{o}/\kindex{D}{o}=0.8$.
For $L/D>4.6$, the positive FTLE ridge starts catching up with the vortex core and negative ridge (\cref{fig:ftle}c) and the stream-wise length rapidly decreases and converges to a value of $\kindex{L}{o}=\SI{1.76}{D}$ for $L/D>6$.
This corresponds to an aspect ratio of $\kindex{L}{o}/\kindex{D}{o}=1.4$.
Based on the velocity field (\cref{fig:vort}) and FTLE snapshots (\cref{fig:ftle}), the merging of the secondary vortex with the primary vortex ring occurs for $5<L/D<6$.
During this time interval, indicated by the shaded region in \cref{fig:viscues}b, the FTLE boundaries contract and push vorticity from the tail of the FTLE bound region towards the primary core line to merge.

To quantify the asymmetry of the FTLE bound area with respect to the vortex core location, we introduce the following asymmetry parameter:
\begin{equation}\label{eq:asymm}
a=\frac{\kindex{L}{p}-\kindex{L}{n}}{\kindex{D}{o}}\quad,
\end{equation}
with \kindex{L}{n} the distance from the stream-wise location of the vortex core to the leading nFTLE intersection, and \kindex{L}{p} the distance from the stream-wise location of the vortex core to the lagging pFTLE intersection with the centreline (\cref{fig:viscues}a).
Values of $a$ close to zero indicate symmetric FTLE boundaries with respect to the vortex core, higher positive values indicate an asymmetric tail heavy FTLE enclosed area.
For $3<L/D<4.5$, the asymmetry increases similarly to the vortex length due to the lagging of the positive FTLE ridge.
For $L/D>4.5$, the positive FTLE ridge catches up with the core line and the asymmetry parameter decreases and reaches zero at $L/D=6.2$.
The FTLE boundaries become fully symmetric with respect to the vortex core after vortex merging.

\subsection{Evolution of integral quantities during vortex merging}
The variation in the vortex ring topology during merging affects the amount of vorticity and its distribution inside the vortex core, which influences the vortex circulation, hydrodynamic impulse, and energy \citep{gharibUniversalTimeScale1998, deguyonScalingTranslationalVelocity2021}.
By analysing the FTLE ridges and the local extrema in the pressure field of vortex rings created by synthetic jets, \cite{lawsonFormationTurbulentVortex2013} demonstrated that vortex rings can continue to grow after pinch-off due to interaction with their environment.
To quantify the evolution of the vortex ring development during the entire formation process, we analyse here the temporal evolution of the integral quantities of the vortex, including its circulation, energy, and resulting translational velocity.

\begin{figure}[t!]
\includegraphics{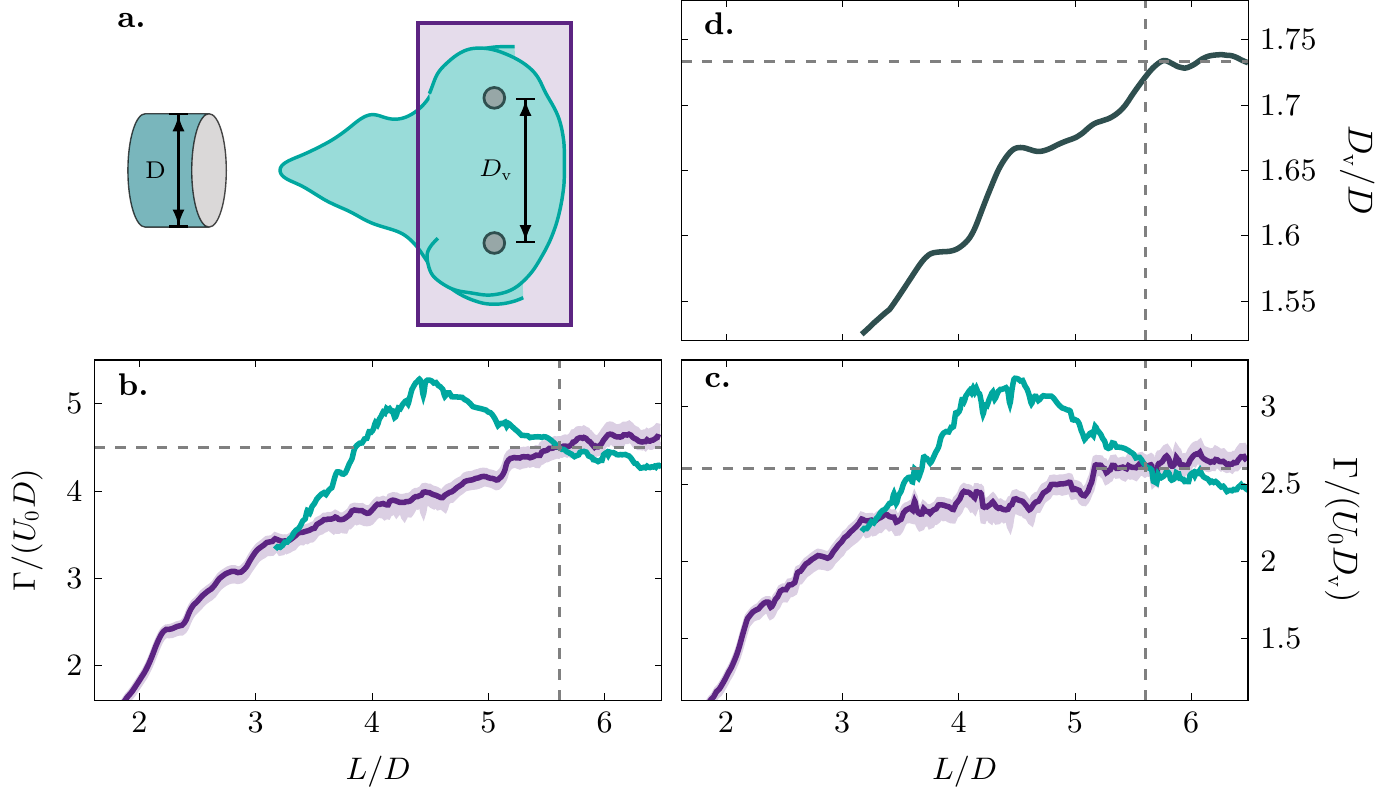}
\caption{Temporal evolution of the vortex circulation and vortex diameter.
Circulation values obtained by integration of the experimentally obtained out-of-plane vorticity within the area bound by the positive and negative FTLE ridges are presented in violet, values obtained by integration within a rectangular box centred around the primary vortex centres are presented in turquoise.
The light blue shaded region represents the uncertainty based on variations in the box height.
(\textbf{a}) Sketch of the two integration areas.
(\textbf{b}) Circulation normalised by \kindex{U}{0} and the nozzle diameter $D$.
(\textbf{c}) Circulation normalised by \kindex{U}{0} and the vortex diameter \kindex{D}{v}.
(\textbf{d}) Evolution of the vortex diameter normalised by the nozzle diameter.
}\label{fig:iqs}
\end{figure}

The circulation of the vortex ring is computed as the surface integral of the average between the positive and negative out-of-plane vorticity:
\begin{equation}
\label{equation:gamma}
\Gamma = \iint\limits_{A} \frac{|\omega|}{2}\, \dd A \quad. 
\end{equation}
We have calculated the circulation within the area bound by the FTLE ridges and within a rectangular area centred around the vortex centre such that it only contains the primary vortex ring (\cref{fig:iqs}a).
The extent of the rectangular box is defined based on the trailing pressure maximum following the procedure presented by \cite{lawsonFormationTurbulentVortex2013}.
The temporal evolution of the non-dimensional circulation in both integration areas is presented in \cref{fig:iqs}b,c.
The circulation is non-dimensionalised by the characteristic velocity $\kindex{U}{0}$ and the nozzle outlet diameter $D$ in \cref{fig:iqs}b and by $\kindex{U}{0}$ and the vortex diameter \kindex{D}{v} in \cref{fig:iqs}c.
The vortex diameter normalised by the nozzle diameter is presented in \cref{fig:iqs}d.

Due to the initial tail-heavy asymmetry of the FTLE boundaries with respect to the vortex core, the circulation inside the FTLE contour is higher than the circulation in the box until $L/D\approx 5.5$.
The width of the FTLE boundary converges to the width of the rectangular contour post vortex merging and the two circulation curves in \cref{fig:iqs}b,c converge to $\Gamma/(\kindex{U}{0}D)\approx\num{4.3}$ and $\Gamma/(\kindex{U}{0}\kindex{D}{v})\approx\num{2.5}$.
The final value of the vortex circulation when normalised by the vortex diameter instead of the nozzle diameter is close to the non-dimensional values reported for propulsive vortex rings generated with a piston cylinder \citep{gharibUniversalTimeScale1998} and drag vortices behind a translating cone \citep{deguyonScalingTranslationalVelocity2021}.

The difference between the circulation in the FTLE contour and in the rectangular contour for $L/D>5.5$ is attributed to vorticity outside the FTLE contour in the radial direction.
For $L/D<5.5$, the difference is attributed to circulation in the trailing shear layer that will eventually be fed into the primary vortex ring through merging.
The maximum circulation in the FTLE boundary in measured when the stream-wise length of the vortex is also maximal (\cref{fig:viscues}b).
The circulation in the rectangular contour gradually increases and does not attain a local maximum prior to converging at $L/D>5.5$.
When the circulation inside the vortex ring is normalised by the vortex diameter instead of the nozzle diameter, the non-dimensional circulation already reaches \SI{90}{\percent} of its final value after $L/D=3.6$, which is well before vortex merging takes place.
The additional volume and associated vorticity that is added to the main vortex ring due to merging does not significantly alter the non-dimensional circulation based on the vortex size, but primarily leads to an increase in the vortex core diameter (\cref{fig:iqs}d).
The increase in the vortex diameter during the merging process is confirmed for vortex rings generated in different configurations, such as orifice-generated vortex rings \citep{limbourgFormationOrificegeneratedVortex2021a}.
After merging, the vortex core diameter converges to $\kindex{D}{v}/D=1.73$ at the same time as the non-dimensional circulation converges.

Based on the evolution of the vortex circulation, we can calculate a vortex formation number.
The formation number of a vortex ring is typically obtained as the stroke ratio at which the circulation in the entire domain equals the steady-state circulation value inside the vortex ring.
Following this convention, we obtain a vortex formation number of \num{3.3} which is within the range of formation numbers observed for propulsive vortex rings generated for different stroke ratios \citep{dabiriOptimalVortexFormation2009}.
Note that the evolution of the total circulation in the domain is not included in the figures as it quickly exceeds the axis range selected for display.

The non-dimensional energy of the vortex ring serves as a measure for the distribution of the vorticity inside the vortex ring \citep{deguyonScalingTranslationalVelocity2021}.
According to \citet{gharibUniversalTimeScale1998}, the non-dimensional energy $E^*$ is defined as:
\begin{equation}
\label{equation:estar}
E^* = \frac{E}{\sqrt{\rho I \Gamma^3}} \quad ,
\end{equation}
with $E$ the kinetic energy, $\rho$ the fluid density, $I$ the impulse, and $\Gamma$ the circulation of the vortex ring.
The kinetic energy is computed as:
\begin{equation}
\label{equation:energy}
E=\pi \rho \iint\limits_{A} \omega \psi \dd A \quad,
\end{equation}
where $\psi$ is the stream function, obtained from integrating the Cauchy Riemann equations for the axisymmetric vortex ring.
The stream function is computed here in the entire domain and integrated within the integration area $A$.
Similar to the procedure applied to compute the circulation (\cref{fig:iqs}a), we consider again two integration areas.
One area is bound by the FTLE ridges, the other one is a rectangular area centred around the vortex centres.
The impulse of the vortex ring is computed as:
\begin{equation}
I = \pi \rho \iint\limits_{A} \omega |r|^2 \dd A \quad,
\end{equation}
where $|r|$ is the distance away from the vortex centreline.

\begin{figure}[tb!]
\includegraphics{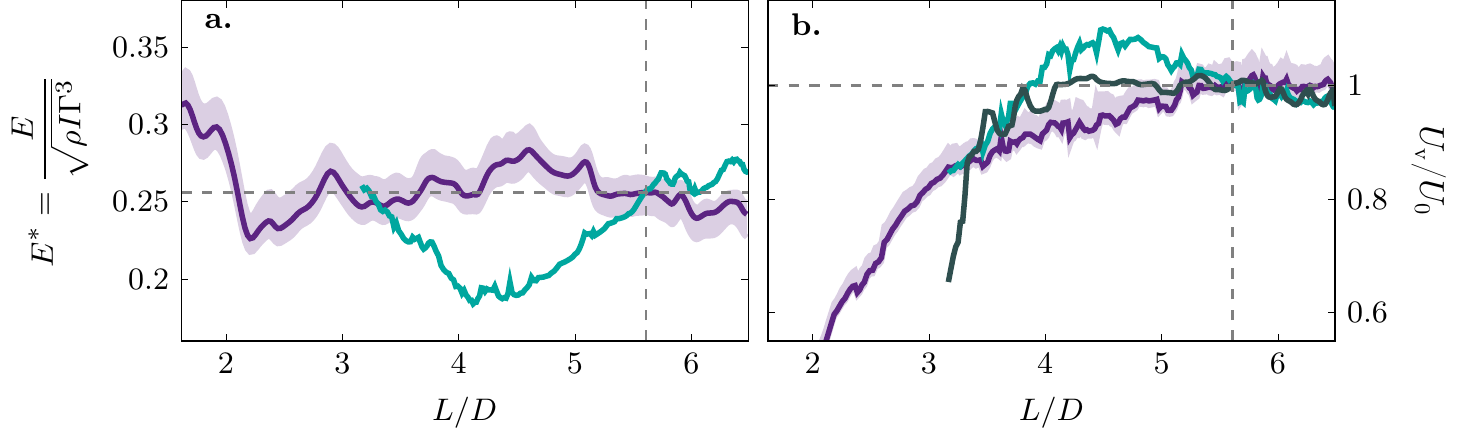}
\caption{Temporal evolution of (\textbf{a}) the non-dimensional energy of the vortex and (\textbf{b}) the translational velocity of the experimentally observed vortex according to \cref{equation:utrans} and based on the tracking of the vortex centre locations.
Values obtained by integration in the area bound by the FTLE ridges are presented in violet, values obtained by integration within a rectangular box centred around the vortex centres are presented in turquoise.
}\label{fig:estar}
\end{figure}

The non-dimensional energy for the two integration areas is presented in \cref{fig:estar}a.
The evolution of $E^*$ within the FTLE boundary is only shown as a reference.
The asymmetric shape of the FTLE ridges for $L/D<6$ indicate that the primary vortex ring is not isolated and we cannot directly interpret the value of the non-dimensional energy bound by the FTLE ridges as a measure of the vorticity distribution within the vortex ring.
The non-dimensional energy inside the rectangular contour that contains only the primary vortex ring drops to a steady state value of \num{0.26} after only three stroke ratios.
This would point again towards a vortex formation number around \num{3}.
The formation number can be interpreted here as the non-dimensional time required for the vorticity to accumulate into a stable distribution but it does not mean that the vortex will not accept additional vorticity or impulse.
After the initial decrease, the non-dimensional energy of \num{0.26} is maintained for the remainder of the vortex merging process while other quantities, such as the circulation and diameter of the vortex ring (\cref{fig:iqs}b-d), continue to evolve.
The continued evolution of the circulation and size of vortex rings after their pinch-off or for formation times beyond the formation number has been observed by \cite{lawsonFormationTurbulentVortex2013, limbourgFormationOrificegeneratedVortex2021a}.

The limit value of the non-dimensional energy observed in this study, \num{0.26}, is lower than the value of \num{0.33} which is typically observed in literature for vortex rings generated with a constant piston or jet velocity \citep{shusserNewModelInviscid1999, nitscheSelfsimilarSheddingVortex2001}.
This lower value is attributed to the kinematics of the vortex propulsor, which influences the velocity and acceleration profile of the jet that feeds the vortex ring \citep{limbourgExtensionUniversalTime2021, kriegApproximatingTranslationalVelocity2013,deguyonEstimatingNondimensionalEnergy2021}.
The nozzle exit velocity decelerates for $t/T>0.2$ up to the end of the compression phase of the vortex generator (\cref{fig:uf}e), which resembles the negative sloping velocity programs implemented by \cite{danailaFormationNumberConfined2018, kruegerSignificanceVortexRing2003}.
The non-dimensional energy of the vortex ring can also be manipulated to values below \num{0.33} by varying the nozzle geometry, such as a converging nozzle or a nozzle with a temporally-varying exit diameter \citep{kriegNewKinematicCriterion2021}.

The additional volume and circulation that will merge with the primary vortex ring for $L/D>3$ does not alter its non-dimensional energy.
This is a great example of the Kelvin-Benjamin variational principle, which states that a vortex will only accept additional vorticity if this does not disturb the vorticity distribution, and thus the non-dimensional energy of the new configuration \citep{benjaminAlliancePracticalAnalytical1976}.
The vorticity level in the trailing jet is high enough to allow it to penetrate into the primary vortex ring.
The primary vortex is able to accept the additional vorticity from the secondary vortex by increasing its radius (\cref{fig:iqs}d) and redistributing the additional vortical fluid such that its non-dimensional energy remains constant (\cref{fig:iqs}a).

The reason the secondary vortex can merge with the primary vortex in the first place is due to their relative translation velocities \citep{maxworthyStructureStabilityVortex1972}.
The vortex translation velocity is calculated from the non-dimensional energy, circulation, and diameter of the vortex ring \citep{saffmanVelocityViscousVortex1970, deguyonScalingTranslationalVelocity2021}:
\begin{equation}
\label{equation:utrans}
\kindex{U}{v} = \frac{\Gamma}{\pi \kindex{D}{v}} \bigg(E^* \sqrt{\pi}+\frac{3}{4} \bigg)\quad.
\end{equation}

\Cref{equation:utrans} is valid for steady vortex rings with a thin core \citep{saffmanVelocityViscousVortex1970}.
An error of $\approx\SI{3}{\percent}$ is obtained for the vortex translational velocity by taking into account the effect of the core thickness, based on the second-order correction proposed by \cite{fraenkelExamplesSteadyVortex1972} and \cite{shusserNewModelInviscid1999} for Norbury vortices \citep{norburySteadyVortexRing1972}.

The temporal evolution of the translational velocity was computed for the FTLE boundary and the rectangular contour enclosing solely the primary vortex and is presented in \cref{fig:estar}b.
The translational velocity based on the tracking of the vortex centres is included in \cref{fig:estar}b for comparison.
The translational velocity of just the primary vortex ring based on the integral quantities rapidly increases during the first three stroke ratios to a value of $\kindex{U}{v}/\kindex{U}{0}=0.8$.
The velocity based on the tracking the core converges to $\kindex{U}{v}/\kindex{U}{0}=1$ after four stroke ratios.
This rapid increase in the translational velocity causes the core to move away from the nozzle exit.
At $L/D=3$, the core has already move one diameter away from the outlet (\cref{fig:ftle}a,e) which makes it harder to directly feed additional vorticity into the primary vortex, but it does not mean that the vortex is not able to accept additional vorticity at a later stage.
The early physical distancing of the primary vortex ring is a direct consequence of the specific time-varying outlet velocity profile of our vortex generator.

The translational velocity based on the integral values computed for the FTLE boundary increases beyond the velocity of the primary vortex for $4<L/D<5.5$.
This difference is attributed to the higher translational velocity of the secondary vortex in the tale of the FTLE bounded area.
The secondary vortex catches up with the primary vortex and they merge.
The vortex diameter, circulation, and translational velocity all converge to new post-merging values for $L/D>5.5$ while the non-dimensional energy remains at its limiting pre-merging value in agreement with the Kelvin-Benjamin variational principle.

\subsection{Fluid entrainment and detrainment during vortex merging}
\begin{figure}[tb!]
\includegraphics{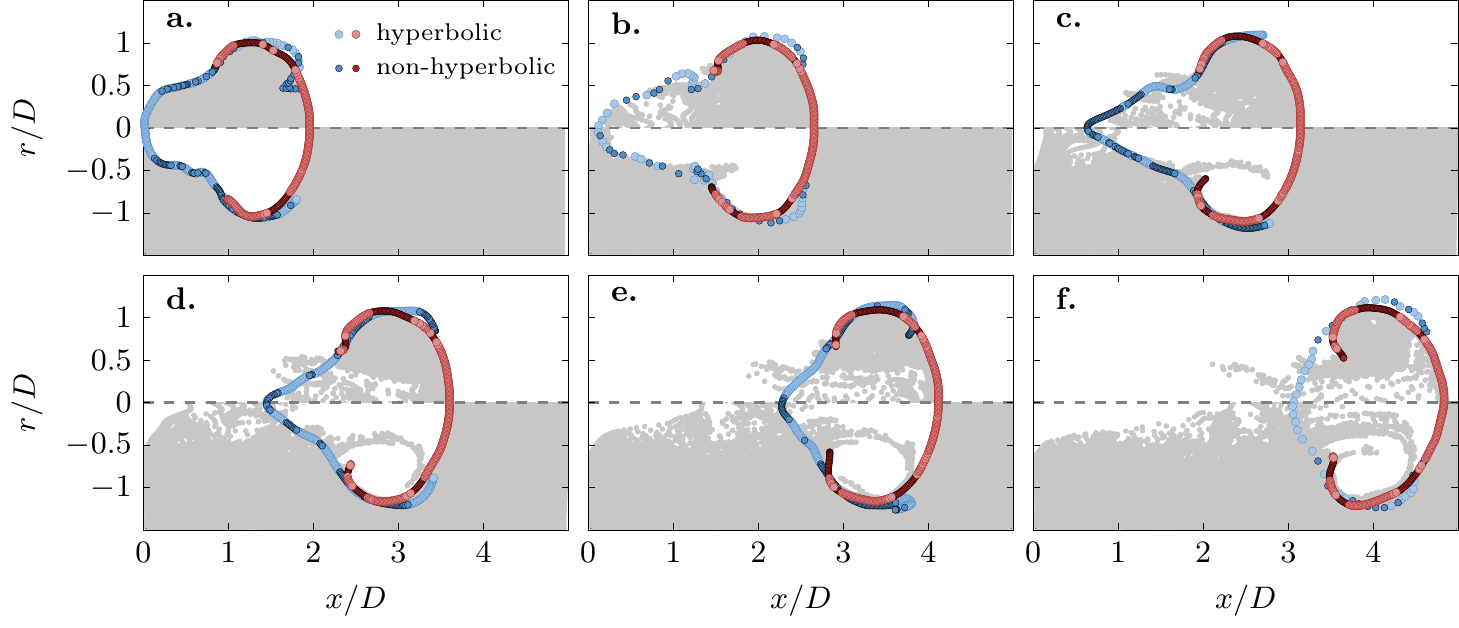}
\caption{
Snapshots of artificial seed particles convected by the experimentally measured flow.
Top half show particles initially inside the FTLE boundaries, bottom half show particles initially outside the FTLE boundaries.
White symbols correspond to points along the FTLE ridges that are non-hyperbolic, filled coloured symbols correspond to points along the FTLE ridges that are hyperbolic.
Initial position of the particles at (\textbf{a}) $L/D$ = \num{3.7}, and their location after convection by the flow at (\textbf{b}) $L/D$ = \num{4.4}, (\textbf{c}) $L/D$ = \num{5.0}, (\textbf{d}) $L/D$ = \num{5.5}, (\textbf{e}) $L/D$ = \num{5.9}, and (\textbf{f}) $L/D$ = \num{6.4}.
} \label{fig:ftraj}
\end{figure}

The entrainment and detrainment of fluid into the merging of the primary and secondary vortex is visualised using a Lagrangian approach.
Artificial seed particles are placed inside and outside the vortex boundaries marked by the FTLE ridges $L/D = 3.7$ and are convected with the flow.
The initial time of particle seeding is selected such that the FTLE ridges demarcating the vortex have fully formed and form a closed contour.
The locations of the seed particles at different time instants during the compression stage of the bulb are presented in \cref{fig:ftraj}.
The top half of the snapshots in \cref{fig:ftraj} present the results for particles that were initially inside the FTLE boundaries and show fluid detrainment.
The bottom half of the snapshots in \cref{fig:ftraj} present the results for particles that were initially outside the FTLE boundaries and show fluid entrainment.

By definition, fluid particles move around the negative ridge and are attracted by the positive FTLE ridge \citep{hallerLagrangianCoherentStructures2002, shaddenTransportStirringInduced2007}.
Occasionally, particles surpass the positive FTLE ridge to enter the vortex boundary (\cref{fig:ftraj}d-f).
The particles leave through the formation of lobes in the negative FTLE ridge, in the process known as tail shedding \citep{shaddenLagrangianAnalysisFluid2006a, deguyonModellingVortexRing2021}.

To understand how particles can cross FTLE ridges we compute the strain rates normal to the ridges.
The strain rates allow us to verify whether the ridges are indeed hyperbolic Lagrangian coherent structures \citep{hallerLagrangianCoherentStructures2002, greenUsingHyperbolicLagrangian2010a}.
The positive FTLE ridge, which repels particles, is a hyperbolic repelling material line if the strain rate normal to the ridge is positive.
A negative FTLE ridge is an attracting material line if the strain rate normal to the ridge is negative \citep{hallerLagrangianStructuresRate2001}.
The sign of the strain rates on the positive and negative FTLE ridges are indicated by the markers in \cref{fig:ftraj}.
The negative FTLE ridge continuously maintains negative strain rates on the ridge throughout the process, confirming that it is a hyperbolic attracting material line \citep{greenDetectionLagrangianCoherent2007a}.
The entrainment of fluid particles into the vortex ring across the positive FTLE ridges occurs where the ridges are locally non-hyperbolic (\cref{fig:ftraj}b-e).
The positive FTLE ridge evolves into an entirely hyperbolic repelling ridge post vortex merging when the ridges symmetrically enclose the primary vortex core (\cref{fig:ftraj}f).

\subsection{Evolution of pressure field during vortex merging}
The Lagrangian analysis provides detailed and accurate insight into the formation and development of the vortex ring generated by our propulsor.
But, the Lagrangian analysis is computationally expensive, requires time resolved flow field data, and is not suitable for in-situ optimisation and control of the time-dependent exit velocity profile.
Local instantaneous pressure measurements would be preferential for this purpose.
To evaluate the potential of pressure-based indicators of vortex formation we need to explore how pressure features in the flow field related to the previously extracted Lagrangian boundaries.

\begin{figure}[tb!]
\includegraphics{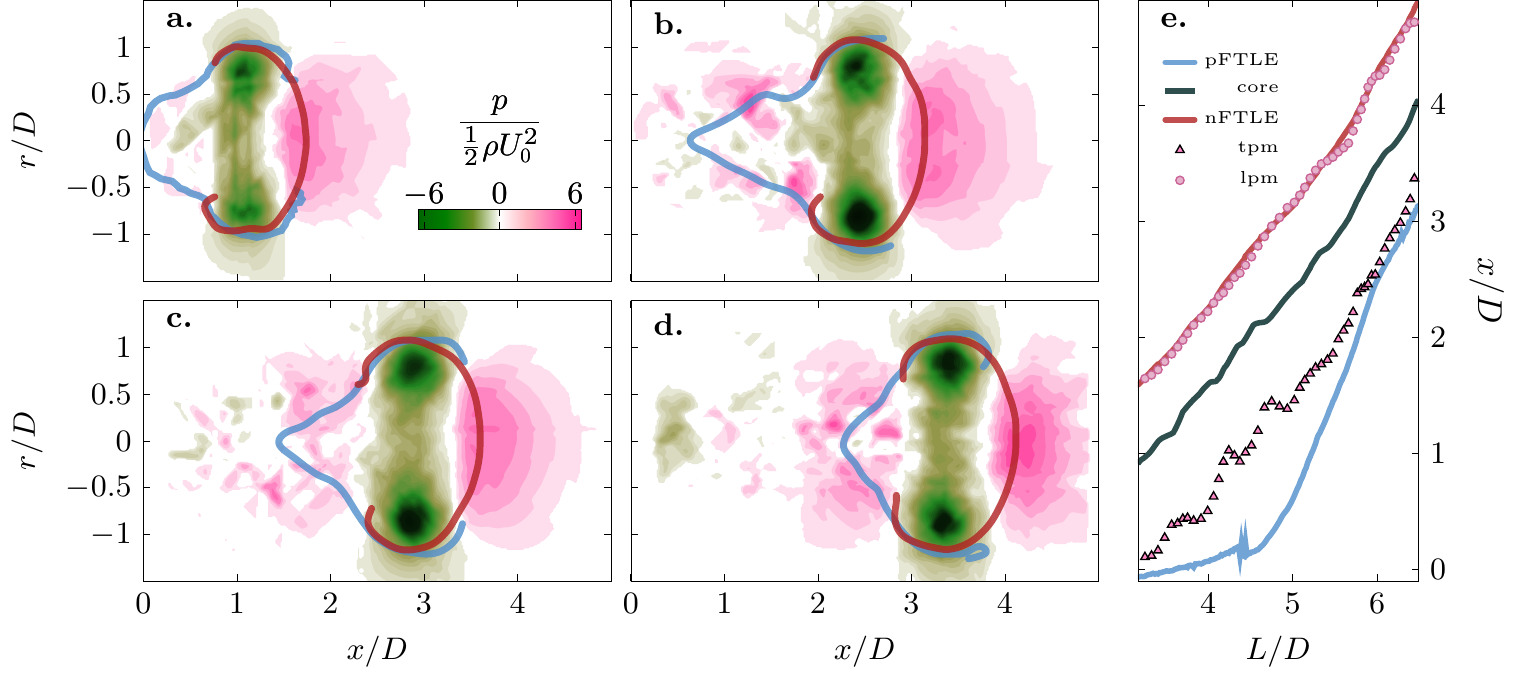}
\caption{Snapshots of the pressure field derived from the experimentally measured velocity field at (\textbf{a}) $L/D$ = \num{3.4}, (\textbf{b}) $L/D$ = \num{5.0}, (\textbf{c}) $L/D$ = \num{5.5}, and (\textbf{d}) $L/D$ = \num{5.9}.
(\textbf{e}) Evolution of the stream-wise location of the vortex core, of the crossing of the positive and negative FTLE ridges with the centreline, and of location of the leading and trailing pressure maxima (lpm, tpm) along the centreline as a function of the stroke ratio.
}\label{fig:press}
\end{figure}

The pressure field is computed from a direct integration of the velocity field \citep{dabiriAlgorithmEstimateUnsteady2014} and presented in \cref{fig:press} for selected snapshots.
The FTLE ridges are added atop the pressure fields for comparison.
The pressure minima reliably indicate the location of the vortex core in all the snapshots.
The pressure minimum becomes stronger with increasing stroke ratio.
A high pressure region called the leading pressure maximum forms ahead of the vortex core and has a local maximum where the negative FTLE ridge intersects with the centreline (\cref{fig:press}a).
Local high pressure regions or trailing pressure maxima, emerge aft of the primary vortex core for $L/D > 3.5$.
These regions are scattered pre vortex merging (\cref{fig:press}b,c) and combine to a single more coherent region centred around the intersection of the positive FTLE ridge with the centreline post merging (\cref{fig:press}d).

The evolution of the stream-wise locations of the vortex core, of the intersection of the positive and negative FTLE ridges with the centreline, and of the location of the leading and trailing pressure maxima along the centreline are summarised in \cref{fig:press}e.
The trajectory of the leading pressure maximum and the negative FTLE ridge coincide perfectly.
The trailing pressure maximum is initially ahead of the positive FTLE ridge which lags behind the primary vortex core.
For $4.4<L/D<6$, the trailing pressure maximum cannot be reliably identified in all snapshots.
We use a linear interpolation to fill the gaps.
The locations of the trailing pressure maximum and the positive FTLE ridge match closely post-merging.

The leading and trailing pressure maxima, like the positive and negative FTLE ridges, reliably indicate the upstream and downstream bounds of the vortex ring \citep{lawsonFormationTurbulentVortex2013}.
The ridges and the pressure maxima symmetrically enclose the primary vortex core and act as physical barriers that prevent additional fluid from entering the vortex ring.
The pressure maxima can serve as reliable observables for vortex ring formation and shedding that can be used for future optimisation and control of the driving jet profile of bio-inspired vehicles and other vortex generating systems.
A pressure-based methodology does not require time-resolved data and is computationally less expensive than calculating the FTLE field.
Our results disclose new possibilities to incorporating pressure sensors and probes in the flow to measure vortex ring properties in-situ.

\section{\label{sec:conclusion}Conclusion}
We have presented here a bio-inspired jet propulsor that combines the body morphologies of two marine organisms, the bell muscle of the jellyfish and the compression kinematic of a bivalve.
Our propulsor generates a non-linear time-varying exit velocity profile by compressing and relaxing a flexible bulb with two rigid arms and has a finite volume capacity.
The formation process of the vortices generated by this jet profile is analysed using time-resolved velocity field measurements.
The temporal evolution of the vortex topology and its integral quantities are analysed based on the finite-time Lyapunov exponent field and the pressure field, both derived from the velocity data.

When fluid is ejected by our vortex propulsor, a coherent vortex ring is formed.
This primary vortex ring rapidly moves away from the nozzle exit during the compression phase and a trailing shear layer with a secondary vortex emerges.
The secondary vortex has a higher translation velocity than the initial primary vortex and both merge before the end of the bulb compression.
Analysis of the temporal evolution of the ridge in the FTLE field during vortex merging reveal that the vortex length increases beyond its diameter pre-merging due to the lagging of the positive FTLE ridge.
During vortex merging, the vortex length contracts and its diameter increases such that additional vorticity is accepted by the primary vortex ring without changing its non-dimensional energy, in agreement with the Kelvin-Benjamin variational principle.
The vortex diameter, circulation, and translational velocity all converge to new post-merging values post-merging.

Our Lagrangian analysis provides detailed and accurate insight into the formation and development of the vortex ring generated by the non-linear evolution of the stoke ratio generated by our propulsor.
However, this type of analysis is computationally expensive, requires time-resolved flow field data, and cannot be conducted in-situ to provide input for optimisation and control of the time-dependent exit velocity profile.
An alternative pressure-based methodology does not require time-resolved data and can be applied on a single snapshot.
We reveal that the trajectories of the pressure maxima that lead and trail the vortex core coincide with the trajectories of the negative and positive FTLE ridges pre and post vortex merging.
During vortex merging, the trailing pressure maximum is less pronounced.
The local pressure maxima can serve as reliable observables for vortex ring formation and shedding that could be picked up by pressure sensors integrated in the surface around the nozzle exit.

Our results provide novel insights into the evolution of integral quantities of vortex rings during merging and can aid and inspire the further design and control of underwater vehicles that uses pulsatile jet propulsion.
Even though our design derives inspiration from biological organisms to create propulsion by periodically produce vortex rings, it does not have the full range of adaptivity displayed by the biological examples.
The vortex rings we generated here are fully axisymmetric, and only provide propulsion in one direction.
Jellyfish and bivalves exploit a variety of both axisymmetric and asymmetric vortex rings to manoeuvre and interact with obstacles in their environments.
The formation and characterisation of asymmetric vortex rings with the vortex propulsor, where three-dimensional effects are involved, will be the subject of future investigations.


\begin{Backmatter}

\paragraph{Funding Statement}
This work was supported by the Swiss National Science Foundation (grant nr. 200021175792).

\paragraph{Declaration of Interests}
The authors declare no conflict of interest.


\paragraph{Data Availability Statement}
The data that support the findings of this study can be made available upon request.


\bibliography{vm_paper}
\bibliographystyle{flow}

\vfill
\newpage
\paragraph{Supplementary Material}
\paragraph{Experimental set-up}
The experimental setup is shown in \cref{fig:expsetup}.
\cref{fig:expsetup}a shows the vortex propulsor ejecting a vortex ring.
\cref{fig:expsetup}b shows the arrangement of the high-speed camera and \ang{45} mirror to image the field of view indicated by the dashed rectangle.
\cref{fig:expsetup}c shows the LED light sheet optics (LED Pulsed System, ILA\_5150 GmbH, Germany).

\begin{figure}[h!]
\centering\includegraphics{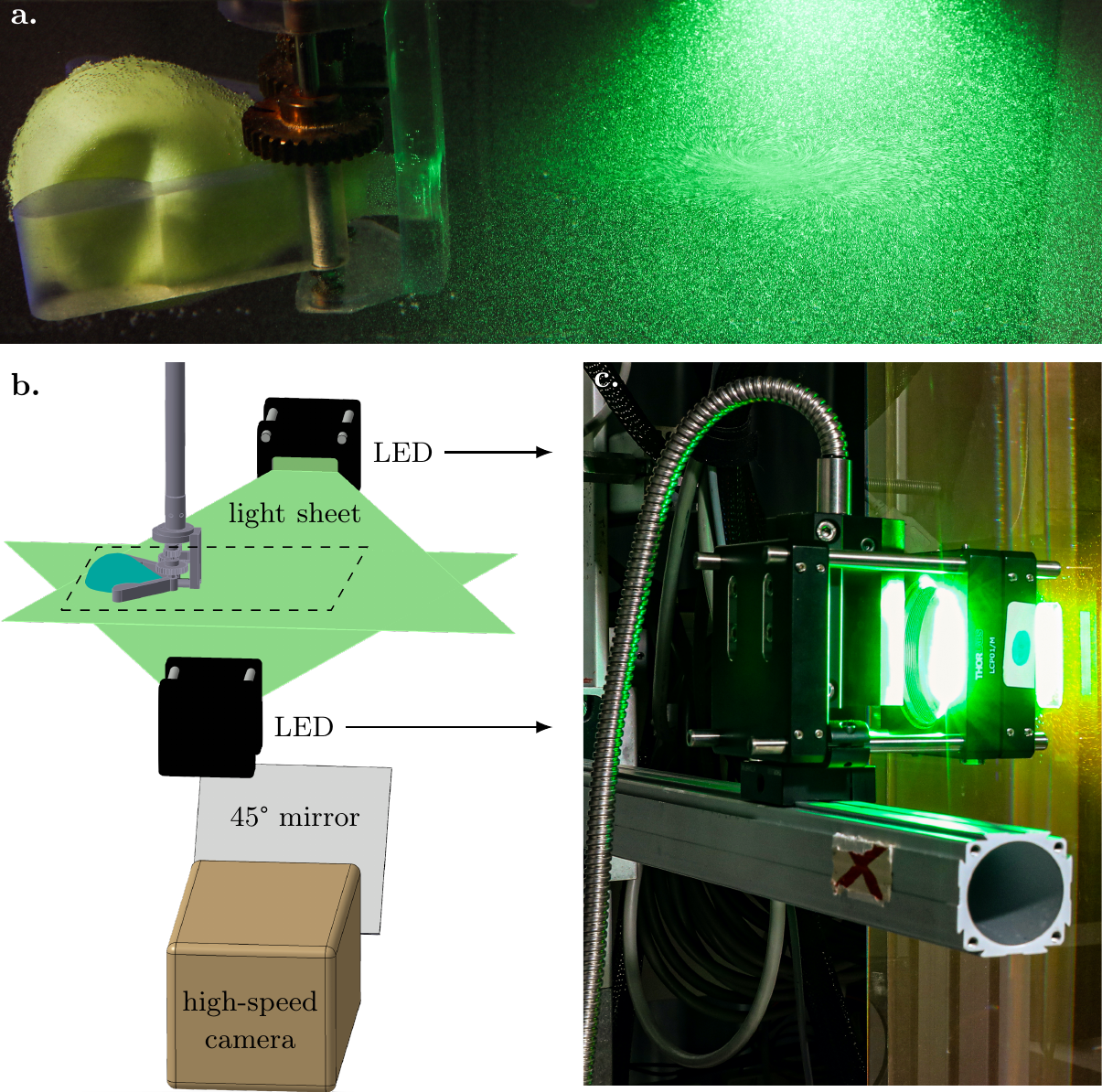}
\caption{(\textbf{a}) Photograph of the compressing propulsor and the ejected vortex ring.
(\textbf{b}) Schematic of the experimental set-up including the arrangement of the high-speed camera, \ang{45} mirror, and light sheet optics with respect to the vortex propulsor.
(\textbf{c}) Photograph of the LED light sheet optics.
}\label{fig:expsetup}
\end{figure}

\paragraph{Streamfunction}
The iso-contours of the stream function calculated from the the two-dimensional velocity field in the symmetry plane are presented in \cref{fig:streamfunc} for two instantaneous snapshots at $L/D=5$ and $L/D=6.4$.
The corresponding FTLE contours demarcating the vortex ring are overlaid for comparison.
The stream function in the vortex frame of reference is given by \citep{fraenkelExamplesSteadyVortex1972}:
\begin{equation}
\kindex{\psi}{v} = \psi - \frac{1}{2}\kindex{U}{v}r^2\quad,
\end{equation}
where $\psi$ is the stream function in the laboratory frame of reference and \kindex{U}{v} is the translational velocity of the vortex ring computed from \cref{equation:utrans}.

\begin{figure}[tb!]
\centering\includegraphics{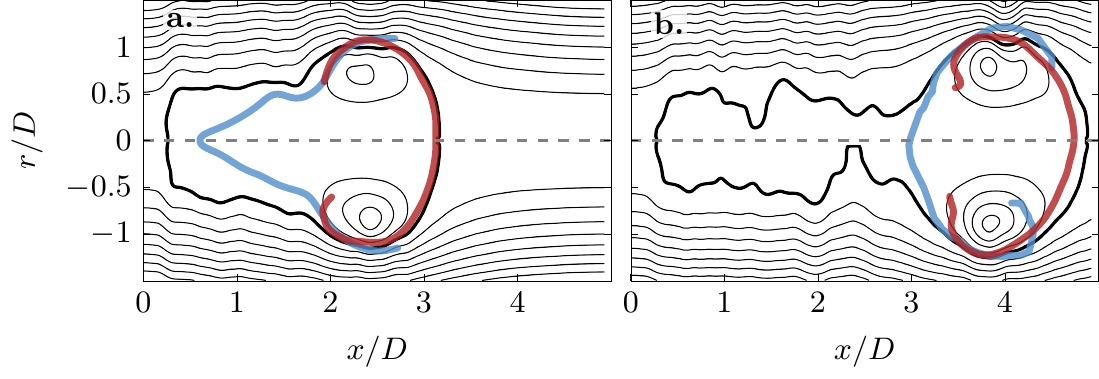}
\caption{Iso-contours of the stream function in the vortex frame of reference at $L/D = 5$ (\textbf{a}) and $L/D = 6.4$ (\textbf{b}).
The contour values range from \num{-8e-5} to \num{8e5} with step size of \num{1e-5}.
The thick black contour is the $\psi=0$ dividing stream line.
The positive (blue) and negative (red) FTLE ridges are superimposed in both images.
}\label{fig:streamfunc}
\end{figure}

The dividing streamline, indicated by the $\psi =0$ contour (thick black line in \cref{fig:streamfunc}), demarcates the vortex.
The negative FTLE ridge is almost perfectly on top of the  $\psi=0$ contour for during vortex formation and only small deviations occur during merging and separation.
The positive FTLE ridge lies inside the $\psi =0$ contour.
The $\psi =0$ contour is not able to distinguish between the starting vortex ring and the trailing jet for the current spatial and temporal resolution.
The FTLE contours are more robust and reliable indicators of the material boundaries of the vortex ring.

The values of the stream function are used to compute the energy of the vortex ring using \cref{equation:energy}, but for the integration contour we used the FTLE boundaries or a rectangular region around the vortex ring centreline.
The extent of the rectangular box is determined based on the trailing pressure maximum following \citet{lawsonFormationTurbulentVortex2013}.

Alternative to \cref{equation:energy}, the energy can be computed for axisymmetric fields with zero swirl by directly integrating the velocity field \citep{limbourgFormationOrificegeneratedVortex2021a, kriegModellingCirculationImpulse2013}:
\begin{equation}
E=\frac{1}{2} \rho \pi \iint\limits_{A} (u^2+v^2)r \dd A \quad.
\end{equation}
The non-dimensional energy obtained from this direct velocity-field integration has a difference of less than \SI{2}{\percent} from non-dimensional energy obtained from the stream function.

\paragraph{Thrust estimation}
An estimate of the propulsive force generated by the bio-inspired device presented here is obtained from a momentum balance directly behind the nozzle exit at $x/D = 0$.
We assume that the propulsive force experienced by the device is proportional to the mass flow rate of the fluid that is ejected.
The average thrust generated over the time of compression is \SI{0.027}{\newton}.
This result is in the range of average thrust values measured by \cite{kruegerSignificanceVortexRing2003} for propulsive vortex rings from a piston-cylinder with varying formation numbers and acceleration profiles.

\paragraph{Suggested location of pressure sensors}
The results presented in this paper disclose new opportunities to use local pressure sensors as input for closed loop control of the temporal exit velocity profile to improve the propulsion efficiency of under water vehicles utilising pulsatile jet propulsion.
Potential locations of the pressure sensors integrated in the surface around the nozzle exit are presented in \cref{fig:psensors}a.
The predicted sensor output based on the pressure field computed from the measured velocity field is presented in \cref{fig:psensors}b.
The pressure signals indicated a clear drop around $L/D=2$ when the primary vortex ring moves away from the nozzle and the non-dimensional energy has converged to its final value.

\begin{figure}[t!]
\includegraphics{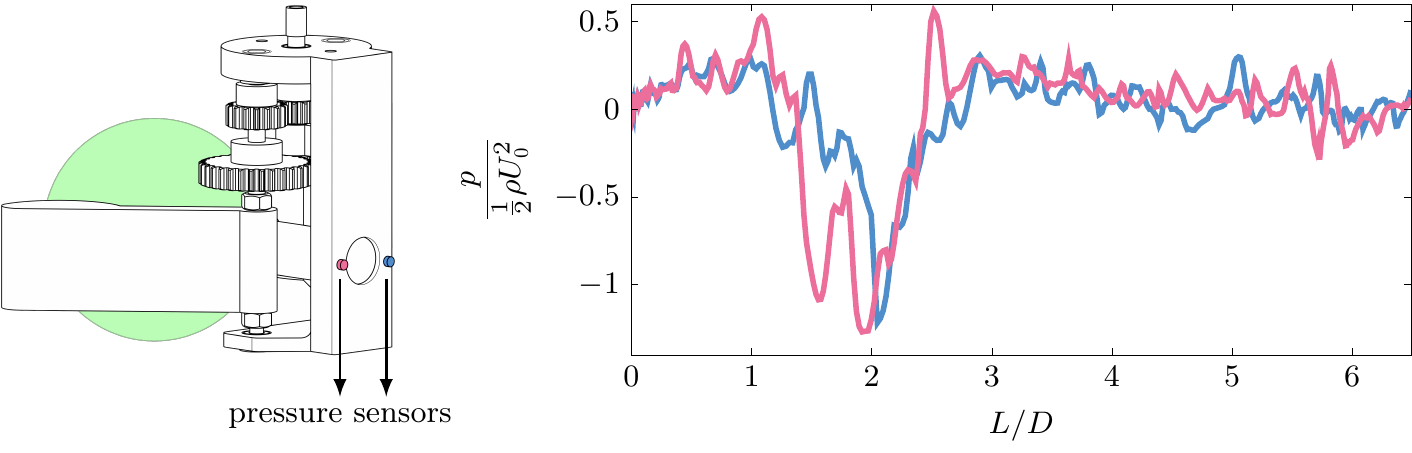}
\caption{(\textbf{a}) Potential locations of integrated pressure sensors to provide input for future closed loop control of the temporal exit velocity profile.
(\textbf{b}) Hypothetical temporal evolution of the pressure signatures captured at these two locations, obtained from the pressure field which was derived directly for the measured velocity field data.
}\label{fig:psensors}
\end{figure}


\end{Backmatter}

\end{document}